\documentclass{kluwer}
\usepackage{epsfig}
\def\HI{H{\sc i}\,}
\begin{document}
\begin{article}
\begin{opening}
\title{Necessary but not Sufficient Conditions to Trigger Starbursts}
\subtitle{CO and HI Observations of Optically-Selected Interacting
Galaxies}
\author{C.  \surname{Horellou},
	R.S.\surname{Booth}, 
	B.  \surname{Karlsson}}
\institute{Onsala Space Observatory, Chalmers University of Technology,
S-439 92 Onsala, Sweden}
\begin{abstract} 
We present the analysis of a survey of atomic and molecular gas
in interacting and merging galaxies 
(Horellou \& Booth  \citeyear{paperI}). 
The sample is optically selected and contains all interacting galaxies
($\approx$ 60 systems) in a well-defined region of the Southern sky
(Bergvall \citeyear{b81}). 
In order to take into account the upper limits due to non-detections, 
we have carried out a survival analysis. The results 
are compared with the ones obtained when ignoring the non-detections. 
We have found evidence for a lower atomic gas content per unit area
in interacting
galaxies compared to isolated ones. 
Except for a few systems with high far-infrared luminosity and correspondingly
high CO fluxes, the interacting galaxies are not unusually CO-bright. 
The observed \HI-deficiency is therefore not due to a conversion from \HI into
H$_2$. Atomic gas from the outer part of the galaxies may have been swept
by tidal interaction, or the optical extent of the galaxies may have 
increased due to the tidal interaction. 
The interacting galaxies seem to be more efficient
at transforming their molecular gas into stars, as indicated by their higher
L$_{\rm{FIR}}$/M(H$_2$) ratio. We found no 
correlation between molecular gas fraction and degree of morphological
distortion. 

\end{abstract}
\end{opening}
\section{Introduction}

Whereas some kinds of galaxy encounters 
(merger of two gas-rich galaxies) are believed to lead
to substantial gas infall and powerful nuclear starbursts, 
others seem to have much more subtle consequences  
for the gas behavior and star-formation properties of the colliders.  
Most studies to date have focussed on the extreme cases rather than on the
interactions themselves, and it is still unclear what the fate of the gas
is in galaxies involved in various types and stages of an interaction.  
In this paper, we study the properties of a complete, optically selected sample
of interacting galaxies in which we have searched for both 
the $^{12}$CO(1--0) and the H{\sc i} lines and estimated the total amounts of
molecular and atomic gas 
(Horellou \& Booth \citeyear{paperI}). 
The sample is not biased towards
far-infrared (hereafter FIR)-luminous galaxies. 
It contains all interacting galaxies in
a well-defined region of the Southern sky with
a blue magnitude brighter than 14.5 $\pm$ 0.3 
(Bergvall \citeyear{b81}).
$UBVRIJHK$ photometry,
optical spectra and images
have been published
(Johansson \& Bergvall \citeyear{jb90},
Bergvall \& Johansson \citeyear{bj95}). 

\section{CO and \HI in an Unbiased Sample}

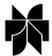
\begin{table}
      \caption[]{Completeness of the dataset}
\begin{tabular*}{\maxfloatwidth}{llllll}
\hline
Parameter      	&FIR$^1$   	&\HI	&$^{12}$CO(1--0)&$D_{25}$$^2$ &B magnitudes$^2$\cr
\hline
Detection rate 	&38/41	 	&26/48	&17/65	&48/48	   &44/55\cr
\hline
\end{tabular*}
\\
\endnotemark{Three galaxies were detected by IRAS at 60 $\mu$m only}\\
\endnotemark{Taken from the NASA Extragalactic Database}
\end{table}
\begin{table}
\def\hf{\hfill}
      \caption[]{Average gas masses and gas surface densities$^3$}
\begin{tabular*}{\maxfloatwidth}{llll}\hline
        log M(H{\sc i})\hf      &log M(H$_2$)\hf & log $\sigma_{\rm{HI}}$\hf
&log $\sigma_{\rm{H_2}}$\hf\\
M$_\odot$	&M$_\odot$   &M$_\odot$ kpc$^{-2}$ &M$_\odot$ kpc$^{-2}$ \\ 
\hline
        9.69$\pm$0.10\hf        &9.36$\pm$0.18\hf       &6.80$\pm$0.10\hf
&6.37$\pm$0.08 (detected)\cr
        9.16$\pm$0.13\hf        &7.85$\pm$0.18\hf       &6.31$\pm$0.10\hf
&5.35$\pm$0.12 (all)\cr
\hline
\end{tabular*}
\\
\endnotemark{
The diameter used here is 
$D_{25}$, the diameter of the blue luminosity at the 25th magnitude per
arcsec$^2$ whereas Haynes \& Giovanelli (1984) use UGC
diameters:
log($D_{UGC}+0.3$)= 1.0173 log($D_{25}$)+0.0519. 
We quote standard errors, whereas they use standard deviations.}
\end{table}
Let us define the gas surface densities  as
$\sigma_{\rm{HI}}$ = M(\HI)/$D^2$ and
$\sigma_{\rm{H_2}}$ = M(H$_2$)/$D^2$, 
where $D$ is the optical diameter of a galaxy. 
Those are not real surface densities since the atomic gas usually extends 
beyond
the optical disk of a galaxy, 
whereas the molecular gas as traced by the CO 
is confined to the inner part of the disk. 
Nevertheless, although being hybrid quantities, 
they provide information on {\sl global} galaxy properties 
and can be compared to
the values determined for other samples.

Table 1 describes the completeness of the dataset.
In order to take into account the non-detections, we have applied
a survival analysis 
(see Isobe et al. \citeyear{isobe}).
We have used the Kaplan-Meyer estimator to 
calculate the average of a dataset containing censored data
(upper limits). 
The mean values of the gas masses in atomic and molecular
form are given in Table 2. We list the results obtained for the detected
galaxies and those estimated for the whole sample. 

\subsection{Comparison with Isolated Galaxies}
{$\bullet$ Interacting Galaxies have less \HI per Unit Area}

We have compared the H{\sc i} surface density of the interacting galaxies 
with those of isolated galaxies observed by Haynes \& Giovannelli 
(\citeyear{hg84}, hereafter HG84), who have derived a canonical value of  
6.81$\pm$0.24  
for $\sigma_{\rm{HI}}$ 
when averaging over all morphological types. 
{\sl The interacting galaxies clearly have a lower HI surface density:} 
6.33$\pm$0.51 for the \HI-detected ones, and 
6.18 for the whole sample 
(here we have taken UGC diameters for comparison with HG84 and not 
$D_{25}$ as in Table II.) The same effect is found when normalizing to the blue
luminosity rather than to the optical area. 

\noindent
{$\bullet$ An Elevated Star Formation Efficiency}
\begin{figure}
\centerline{\epsfig{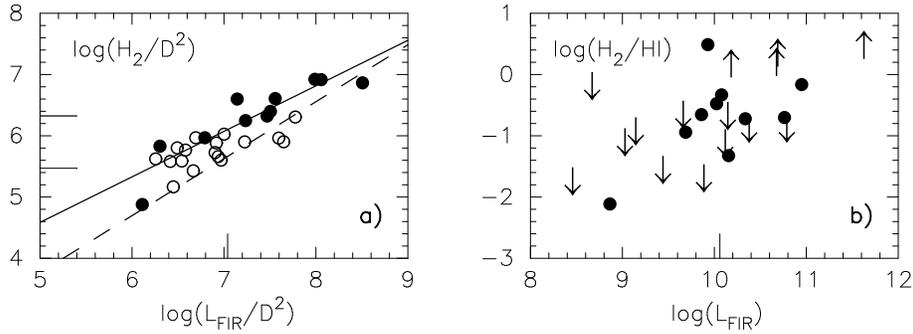}}
\caption{
{\bf a)} The H$_2$ surface density is best correlated with the FIR
surface brightness $\sigma_{\rm{FIR}}$=$L_{FIR}/D_{25}^2$. 
The black dots are the CO-detections,
the open circles the upper limits. The solid line shows the 
linear regression to the detected points, whereas the dashed line is the
result of the survival analysis. The ticks on the axes show the mean
values of the corresponding quantities obtained  
when ignoring the non-detections
and when including them in the survival analysis. 
{\bf b)} The M(H$_2$)/M(\HI) ratio increases with FIR luminosity. 
The upwards pointing arrows represent the galaxies for which
we have upper limits on the \HI mass, whereas the downwards 
pointing arrows represent those with upper limits on the  
H$_2$ mass. 
}
\end{figure}

Using observations of isolated galaxies and of galaxies in the outer parts
of clusters, Casoli et al. (\citeyear{casoli98}) have established a 
linear relationship between $\sigma_{\rm{H_2}}$ and $\sigma_{\rm{FIR}}$.
Their fit is in very good agreement with the one that we obtain for 
the CO-detected interacting galaxies
(solid line in the 
$\sigma_{\rm{H_2}}$-$\sigma_{\rm{FIR}}$ plot displayed on Figure 1a). 
However, if we include the non-detections in a survival analysis, 
our fit falls well below the one for isolated galaxies.  
This means that for the 
same FIR luminosity, interacting galaxies have less molecular gas. 
If one interprets the $L_{FIR}$/M(H$_2$) ratio as an indicator of the
star-formation efficiency (SFE), then 
{\sl interacting galaxies have a higher
SFE compared to isolated galaxies}. 

\subsection{Molecular Gas Fraction and Strength of Interaction}

As pointed out by other authors (e.g., Mirabel \& Sanders \citeyear{mirabel89}),
we find that the fraction of molecular gas in interacting galaxies
increases with the FIR luminosity (Figure 1b). 
We have searched for relations between the gas content or 
tracers of the star formation activity (optical colors, 
FIR luminosity and dust temperature) and the apparent 
degree of morphological distortion, using 
Bergvall's (1981) classification 
(0: undisturbed; 1: weak; 2: medium; 3: strong).
The most strongly disturbed galaxies
do not have significantly different optical colors.
Their dust temperatures  
are slightly higher than than those of the less disturbed ones
but the difference is at a few $\sigma$ level.

\section{Conclusion}
Although all the sample galaxies are in the state of an interaction or
even of a merger, 
their molecular gas fraction and their star-formation activity 
are not significantly 
enhanced.
This can be understood in the light of recent simulations, which clearly
show that 
the evolution of an interacting system is affected by several factors, such as:
the initial conditions (relative mass and 
morphological types of the colliders),
the geometry of the encounter (planar or inclined, prograde or not;
e.g., Howard et al. \citeyear{howard})
and the time scale. 
The amount of molecular gas in particular depends critically on the time, 
since a starburst of $\simeq$ 100 M$_\odot$/yr will exhaust its supply 
in a few 10$^7$ yrs. 
A high concentration of molecular gas 
will be observable
only during that short period. 
Selecting galaxies on the basis of their emission-line or far-infrared
properties introduces a bias toward a particular kind of interaction and/or
stage of merger and excludes the more ``quiet" collisions.  

\begin{acknowledgements}
The authors are grateful to  
the referee, N. Bergvall, and to J.H. Black and F. Casoli 
for pertinent comments. 
C.H. acknowledges financial support from  
the Anglo American Chairman's Fund and SASOL. 
\end{acknowledgements}
{}
\end{article}
\end{document}